\documentclass[structabstract]{raa_rb}
\usepackage{booktabs}
\usepackage{graphicx,times}
\usepackage{natbib,amssymb}
\usepackage{epsfig}
\usepackage{float}
\usepackage{amsmath}
\usepackage{ulem}

\usepackage{mathtools}

\providecommand{\bm}[1]{\mbox{\boldmath$#1$\unboldmath}}

\begin{document}
   \title{He-shell flashes on the surface of oxygen-neon white dwarfs}

   \volnopage{ {\bf 2020} Vol.\ {\bf X} No. {\bf XX}, 000--000}
   \setcounter{page}{1}

   \author{Y. Guo \inst{1,2,3,4},
	       D. Liu \inst{1,2,3,4} ,
	       C. Wu \inst{5}
          \and
               B. Wang \inst{1,2,3,4}
          }

   \institute{Yunnan Observatories, Chinese Academy of Sciences, Kunming 650216, China;
          {\it liudongdong@ynao.ac.cn; wangbo@ynao.ac.cn}\\
          \and
          Key Laboratory for the Structure and Evolution of Celestial Objects, Chinese Academy of Sciences, Kunming 650216, China
          \and
          University of Chinese Academy of Sciences, Beijing 100049, China\\
          \and
          Center for Astronomical Mega-Science, Chinese Academy of Sciences, Beijing 100012, China\\
          \and
          Physics Department and Tsinghua Center for Astrophysics (THCA), Tsinghua University, Beijing 100084, China\\
              }

   \date{Received ; accepted}

\abstract {Accretion induced collapse (AIC) may be responsible for the formation of some interesting neutron star binaries, e.g., millisecond pulsars, intermediate-mass binary pulsars, etc.
It has been suggested that oxygen-neon white dwarfs (ONe WDs) can increase their mass to the Chandrasekhar limit by multiple He-shell flashes, 
leading to AIC events.
However, the properties of He-shell flashes on the surface of ONe WDs are still not well understood.
In this article, we aim to study He-shell flashes on the surface of ONe WDs in a systematic approach.
We investigated the long-term evolution of ONe WDs accreting He-rich material with various constant mass-accretion rates by time-dependent calculations
with the stellar evolution code Modules for Experiments in Stellar Astrophysics (MESA), in which the initial ONe WD masses range from 1.1 to 1.35\,${M}_\odot$.
We found that the mass-retention efficiency increases with the ONe WD mass and the mass-accretion rate,
whereas both the nova cycle duration and the ignition mass decrease with the ONe WD mass and the mass-accretion rate.
We also present the nuclear products in different accretion scenarios.
The results presented in this article can be used in the future binary population synthesis studies of AIC events.
 }

\titlerunning{He-shell flashes on the surface of ONe WDs}

\authorrunning{Y. Guo et al.}

\maketitle


\section{Introduction} \label{1. Introduction}
It has been widely accepted that carbon-oxygen white dwarfs (CO WDs) can form type Ia supernovae
when they grow in mass close to the Chandrasekhar limit ($M_{\rm Ch}$) (e.g., \citealt{1997Sci...276.1378N}, \citealt{2000ARA&A..38..191H}).
Nevertheless, oxygen-neon white dwarfs (ONe WDs)
may undergo accretion-induced collapse (AIC) when their mass approach $M_{\rm Ch}$,
forming neutron stars (NSs) eventually (e.g., \citealt{1980PASJ...32..303M, 1984ApJ...277..791N, 1991ApJ...367L..19N, 2019ApJ...886...22Z}).
AIC events are predicted to be very faint and fast transients, which have not been identified directly so far (e.g., \citealt{1992ApJ...391..228W, 2006ApJ...644.1063D}).
However, it is notable that AIC process can form some interesting objects in the observations, such as X-ray binaries, millisecond pulsars and low-/intermediate-mass binary pulsars (e.g., \citealt{1990ApJ...353..159B, 1991PhR...203....1B, 1992Natur.357..472U, 2002ApJ...565.1107P, 2013A&A...558A..39T, 2015ApJ...800...98A, 2017ApJ...851...58L, 2018MNRAS.477..384L}).
It has been suggested that AIC process may also relate to the r-process nucleosynthesis, the formation of rapidly spinning magnetars and ultra-high-energy cosmic rays like gamma-ray bursts (e.g., \citealt{1998ApJ...493L.101W, 1999ApJ...516..892F, 2007PhR...442..237Q, 1992ApJ...388..164D, 1992Natur.357..472U, 2016ApJ...830L..38M, 2016ApJ...826...97P, 2017arXiv170902221L}).
Furthermore, AIC is a potential source of gravitational wave emission (e.g., \citealt{2010PhRvD..81d4012A}).
For a recent review on the AIC of ONe WDs, see \cite{2020arXiv200501880W}.

Although the AIC process is a non-negligible channel to form NSs, their progenitor model is still unclear.
There are two classic progenitor models of AIC events, that is, the single-degenerate (SD) model and the double-degenerate (DD) model.
In the SD model, an ONe WD accretes H/He-rich material from a non-degenerate companion.
The companion can be a main-sequence (MS) star, a red giant (RG) star or a He star (e.g., \citealt{1990ApJ...356L..51C, 1991ApJ...367L..19N, 2013A&A...558A..39T, 2017ApJ...834L...9B, 2018MNRAS.477..384L, 2018MNRAS.481..439W}).
The accreted material will burn into He first, and then into C and O, and then Ne for H-accreting WDs, or burn into C and O first, and then Ne for He-accreting WDs.
If ONe WD can increase its mass close to $M_{\rm Ch}$, it will collapse into a NS.
Note that the final fate of accreting ONe WDs would not be affected by Urca-process
(e.g., \citealt{2017MNRAS.472.3390S}).
Recent studies suggested that He accretion onto CO WDs may also form NSs owing to the off-center carbon ignition when the mass-accretion rate is above a critical value (see \citealt{2016ApJ...821...28B, 2017MNRAS.472.1593W}).
The DD model involves the merger of two WDs shrinkage of the orbit induced by the gravitational wave radiation,
including the merging of double CO WDs, double ONe WDs, or the merging of an ONe WD with a CO WD (e.g., \citealt{1985ApJ...297..531N, 2004ApJ...615..444S, 2016MNRAS.463.3461S, 2020MNRAS.494.3422L}).

It has been proposed that ONe WD+He star systems could experience AIC and then be recycled to form binary pulsars (e.g., \citealt{2013A&A...558A..39T}).
\cite{2018MNRAS.481..439W} suggested that the ONe WD+He star channel is a dominant channel to produce AIC events in the SD model,
with rates of 0.1$-$0.7$\times$$10^{-3}\rm yr^{-1}$ and their delay times range from 30 to 180\,Myr.
In addition, it has been suggested that the ONe WD+He star channel may be responsible for the formation of observed intermediate-mass binary pulsars with short orbital periods (\citealt{2018MNRAS.477..384L}). 
Moreover, PSR J1802-2124 is a binary containing a 1.24$\pm$0.11\,${M}_\odot$ NS and a 0.78$\pm$0.04\,${M}_\odot$ CO WD companion with an orbital period of 0.7\,d and a spin period of 1.2\,ms, which was first detected by the Parkes Multibeam Pulsar Survey (\citealt{2004MNRAS.355..147F}).
\cite{2018MNRAS.477..384L} suggested that a 1.3\,${M}_\odot$ ONe WD and a 1.0\,${M}_\odot$ He star with an orbital period of 0.47\,d is a possible channel to form systems like PSR J1802-2124.
Note that NS+He star systems may also contribute to some of intermediate-mass binary pulsars with short orbital periods (e.g., \citealt{2013MNRAS.432L..75C}).

He-accreting ONe WDs may also relate to the formation of some other peculiar observed objects,
such as AM Canum venaticorum (AM CVn) binaries and He novae.
AM CVn consists of a CO/ONe WD and a He WD with accretion rates ranging from $10^{-13}$ to $10^{-5}{M}_\odot \rm yr^{-1}$ and extreme short orbital periods in the range of 5$-$65\,min (e.g., \citealt{2001A&A...368..939N, 2009ApJ...699.1365S, 2015ApJ...807...74B}).
There are some WD+He star systems in the observations, e.g., CD-30$^{\circ}$\,11223, KPD 1930+2752, V445 Puppis,
and HD 49798 with its compact companion (e.g., \citealt{2007A&A...464..299G, 2008ApJ...684.1366K, 2009ApJ...706..738W, 2010RAA....10..681W, 2015RAA....15.1813L, 2015A&A...584A..37W, 2018RAA....18...49W}).
In these binaries, V445 Puppis is the only He nova that has been identified in late 2000 by Kanatsu (e.g., \citealt{2000IAUC.7552....1K}).

It has been thought that the mass-accretion process plays an important role in studying binary evolution and the final fate of accreting WD (e.g., \citealt{1982ApJ...253..798N, 2014MNRAS.445.3239P, 2015A&A...584A..37W}).
There exist three regions for different mass-accretion rates:
at high accretion rates, the WD will become a He-giant-like star since the accreted He-rich matter piles up on the surface of the WD (e.g., \citealt{1982ApJ...253..798N});
at low accretion rates, He-shell flashes are triggered due to the unsteady He burning on the surface of the WD (e.g., \citealt{2014MNRAS.445.3239P, 2016ApJ...819..168H});
the accretion rates for steady He-shell burning are limited to a narrow regions (see \citealt{1982ApJ...253..798N, 2015A&A...584A..37W}).
The accretion rate that can trigger He-shell flashes is generally considered to be less than $10^{-6}\,{M}_\odot\,\rm yr^{-1}$ (e.g., \citealt{2004ApJ...613L.129K, 2015A&A...584A..37W}).

The mass-retention efficiency during He-shell flashes has a great influence on binary evolution,
and is defined as the ratio of the processed matter remaining after one cycle of He-shell flash to the accreted mass (e.g., \citealt{2004ApJ...613L.129K}).
Previous studies mainly focus on the He-shell flashes on the surface of CO WDs (e.g., \citealt{1999ApJ...513L..41K, 2004ApJ...613L.129K, 2015A&A...584A..37W, 2017A&A...604A..31W}).
In this work, we aim to investigate multicycle He-shell flashes on the surface of ONe WDs and provide some features of He-accreting ONe WDs, such as the mass-retention efficiencies, nova cycle durations and ignition masses with different accretion rates for various ONe WD masses. We also present the nuclear products in different accretion scenarios. In Sect. 2, we show the basic assumptions and methods in our numerical calculations. We show the numerical results in Sect. 3. Finally, we present discussions and summary in Sect. 4.

\section{Methods}
We use the stellar evolution code Modules for Experiments in Stellar Astrophysics (MESA, version 10398; see \citealt{2011ApJS..192....3P, 2013ApJS..208....4P, 2015ApJS..220...15P, 2018ApJS..234...34P}) to simulate the long-term evolution of He-acrreting ONe WDs. 
Using the suite case $\rm make\_o\_ne\_wd$ in MESA, we evolve pre-MS stars with different masses to the WD cooling phase,
and finally obtain ONe WDs with masses of 1.1, 1.2, 1.3, 1.35\,$M_\odot$. The suite case $\rm wd2$ is used to calculate the accretion WDs and the mass ejection processes during nova outbursts.
The mass-accretion rates are assumed in the range of $3 \times 10^{-7}$$-$$2 \times 10^{-6}$ ${M}_\odot \rm yr^{-1}$, for which the WDs would undergo multicycle He-shell flashes. We set the mass fraction of He and metallicity in the accreted material to be 0.98 and 0.02, respectively.

The MESA nuclear reaction network $\rm co\_burn.net$ is adopted in our calculations, including helium, carbon, oxygen burning coupled by 57 reactions (e.g., $\prescript{3}{}{\mathbf{He}}$,$\prescript{4}{}{\mathbf{He}}$,$\prescript{7}{}{\mathbf{Li}}$,$\prescript{7}{}{\mathbf{Be}}$,	$\prescript{8}{}{\mathbf{B}}$,$\prescript{12}{}{\mathbf{C}}$,$\prescript{14}{}{\mathbf{N}}$,$\prescript{15}{}{\mathbf{N}}$,$\prescript{16}{}{\mathbf{O}}$,$\prescript{19}{}{\mathbf{F}}$,$\prescript{20}{}{\mathbf{Ne}}$,$\prescript{23}{}{\mathbf{Na}}$,$\prescript{24}{}{\mathbf{Mg}}$,$\prescript{27}{}{\mathbf{Al}}$,$\prescript{28}{}{\mathbf{Si}}$).
This nuclear reaction network mainly includes reactions as follows:
	\begin{equation}
	\centerline{ $\prescript{4}{}{\mathbf{He}} + \prescript{4}{}{\mathbf{He}} + \prescript{4}{}{\mathbf{He}} \rightarrow \prescript{12}{}{\mathbf{C}} + \gamma$,}
	\end{equation}
	\begin{equation}
	\centerline{ $\prescript{12}{}{\mathbf{C}} + \prescript{4}{}{\mathbf{He}} \rightarrow \prescript{16}{}{\mathbf{O}} + \gamma$,}
	\end{equation}
	\begin{equation}
	\centerline{ $\prescript{12}{}{\mathbf{C}} + \prescript{12}{}{\mathbf{C}} \rightarrow \prescript{20}{}{\mathbf{Ne}} + \prescript{4}{}{\mathbf{He}}$,}
	\end{equation}
	\begin{equation}
	\centerline{ $\prescript{16}{}{\mathbf{O}} + \prescript{4}{}{\mathbf{He}} \rightarrow \prescript{20}{}{\mathbf{Ne}} + \gamma$,}
	\end{equation}
	\begin{equation}
	\centerline{ $\prescript{20}{}{\mathbf{Ne}} + \prescript{4}{}{\mathbf{He}} \rightarrow \prescript{24}{}{\mathbf{Mg}} + \gamma$,}
	\end{equation}
	\begin{equation}
	\centerline{ $\prescript{12}{}{\mathbf{C}} + \prescript{16}{}{\mathbf{O}} \rightarrow \prescript{28}{}{\mathbf{Si}}$,}
	\end{equation}

In our simulation, the default OPAL opacity is adopted, which is applicable to extra carbon and oxygen during the He-shell flash (e.g., \citealt{1996ApJ...464..943I}).
We set the factor that determines the optical depth ($\tau$) of the outer edge of the stellar model to be 50, i.e., assuming that $\tau$ at the boundary of the stellar model is $50 \times 2/3$.
In addition, the value of mixing-length parameter is set to 2.0, which is the ratio of the mixing length to the pressure scaleheight.
Furthermore, we do not consider the influence of hydrodynamics, rotation and convective overshooting in this work.

In the present work, the super-Eddington wind is adopted when ONe WDs undergo multicycle He-shell flashes.
The Eddington luminosity ($L_{\rm Edd}$) and the wind mass-loss rate ($\dot M$) can be expressed as follows:
\begin{equation}
\centerline{ $L_{\rm Edd} = \frac{4\pi GcM_{\rm WD}}{\tau}$,}
\end{equation}
\begin{equation}
\centerline{ $\dot M = \frac {2\eta_{\rm Edd}(L_{\rm eff}-L_{\rm Edd})}{\upsilon_{\rm esc}^2}$,}
\end{equation} 	
where $c, \tau$, $G$, $\upsilon_{\rm esc}$ are the vacuum speed of light, Rosseland mean opacity,
gravitational constant and the escape velocity at the photosphere of accreting ONe WDs, respectively.
We assume that all energy above the Eddington luminosity is used to eject matter (i.e., the super-Eddington wind factor $\eta_{\rm Edd}$ = 1).
In this case, some of the accumulated material will be blown away when the super-Eddington wind is triggered.
The mass-retention efficiency ($\eta_{\rm He}$) can be expressed as follow:
\begin{equation}
\centerline{$$ $\eta_{\rm He} = \frac {M_{\rm acc}-M_{\rm ej}}{M_{\rm acc}}$ ,$$}
\end{equation}
where $M_{\rm acc}$ is the mass accreted by a ONe WD during a single nova cycle (i.e., from the end of one outburst to the end of the next one) and $M_{\rm ej}$ is the ejected mass. In the present work, we simulate about 20 of successive He-shell flashes and provide the average mass-retention efficiency in the results.

\section{Numerical results}
\subsection{An example of He-shell flashes}
\begin{figure}
	\begin{center}
		\epsfig{file=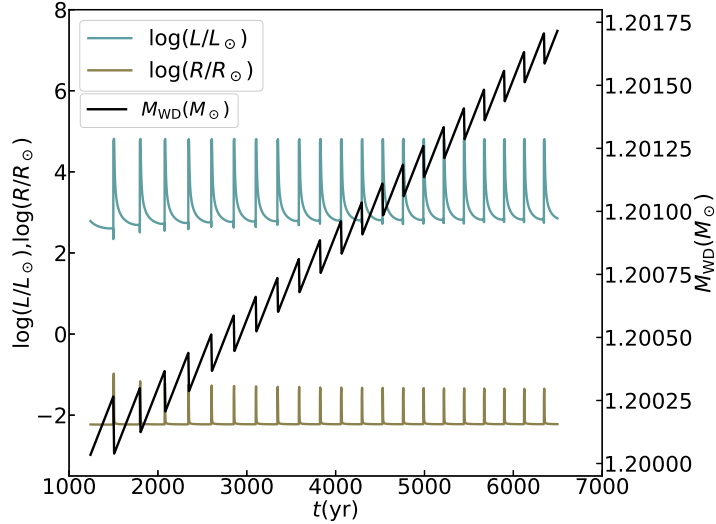,angle=0,width=10.5cm }
		\caption{A representative example of multicycle He-shell flashes on the surface of a 1.2\,$M_\odot$ ONe WD, in which we set the mass-accretion rate ($\dot M_{\rm acc}$) to be 9 $\times 10^{-7}{M}_\odot\,\rm yr^{-1}$.}
	\end{center}
\end{figure}

The He-shell accreted onto the surface of ONe WDs are expected to be ignited and start to burn when its mass reaches a critical value.
Fig.\,1 shows the long-term evolution of the radius, mass and luminosity of a 1.2\,$M_\odot$ ONe WD accreting He-rich material at the mass-accretion rate of $9 \times 10^{-7}$ $M_\odot \rm yr^{-1}$.
From this figure, we can see that some of the accreted material is blown away during He-shell flashes,
and the remaining material leads to the increase of the WD mass.
In this case, the mass of ONe WD would increase to $M_{\rm Ch}$ after much longer evolution, which may lead to the occurrence of the electron-capture reactions of $\prescript{24}{}{\mathbf{Mg}}$ ($\prescript{24}{}{\mathbf{Mg}} \rightarrow \prescript{24}{}{\mathbf{Ne}}$) and $\prescript{20}{}{\mathbf{Ne}}$ ($\prescript{20}{}{\mathbf{Ne}} \rightarrow \prescript{20}{}{\mathbf{O}}$) (e.g., \citealt{1980PASJ...32..303M, 1991ApJ...367L..19N, 2015MNRAS.453.1910S, 2018RAA....18...36W}).
In addition, we can also see the He shell outburst periodically. Especially, the He-shell outbursts tend to become milder as the surface of ONe WD is heated up by the first flash.

\begin{figure}
	\begin{center}
		\epsfig{file=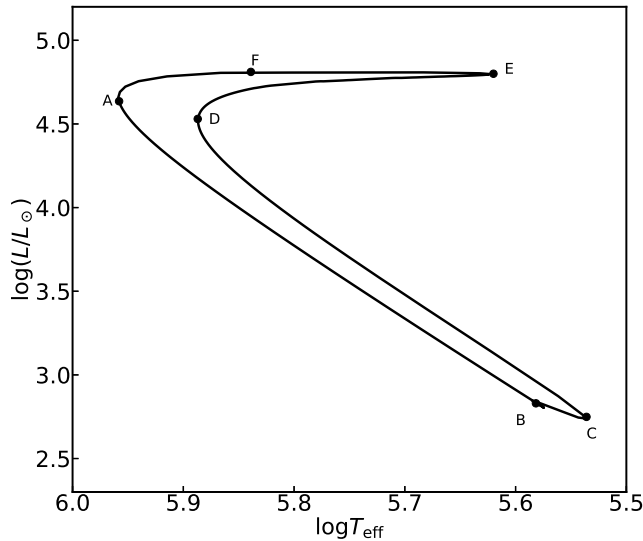,angle=0,width=9.5cm}
		\caption{Hertzsprung-Russell diagram showing a single cycle He-shell flash, with $M_{\rm WD}$ and $\dot M_{\rm acc}$ as 1.2\,${M}_\odot$ and $9 \times 10^{-7}{M}_\odot\,\rm yr^{-1}$, respectively.}
	\end{center}
\end{figure}

Fig.\,2 presents the Hertzsprung-Russell diagram of one cycle of He-shell flashes,
in which $M_{\rm WD}$=1.2\,${M}_\odot$ and $\dot M_{\rm acc}$=$9 \times 10^{-7}{M}_\odot\,\rm yr^{-1}$.
Important phases during the evolution are marked in this figure.
A　$\rightarrow$　B: ONe WD accretes He-rich material, during which the effective temperature $T_{\rm eff}$ and the luminosity $L$ decrease since almost no He-shell burning in this stage to provide the energy to balance the radiative losses;
B　$\rightarrow$　C: the convective shell begin to attain the surface of ONe WD gradually;
C$\rightarrow$　D: $T_{\rm eff}$ and $L$ increase rapidly since the He-shell flash turns into a thermonuclear runaway;
D　$\rightarrow$　E: after the luminosity of ONe WD exceeds $L_{\rm Edd}$, the radius of ONe WD will expand and cause mass loss.
The radius of ONe WD reaches its maximum value at point E;
E　$\rightarrow$　F: radius contraction makes a main contribution to provide energy for mass ejection. Meanwhile, $T_{\rm eff}$ increases and reaches point F where the wind mass loss stops (see also \citealt{1994ApJ...437..802K, 2019MNRAS.490.1678C}).

\subsection{He-shell flashes in different accretion scenarios}
\begin{figure}
	\begin{center}
		\epsfig{file=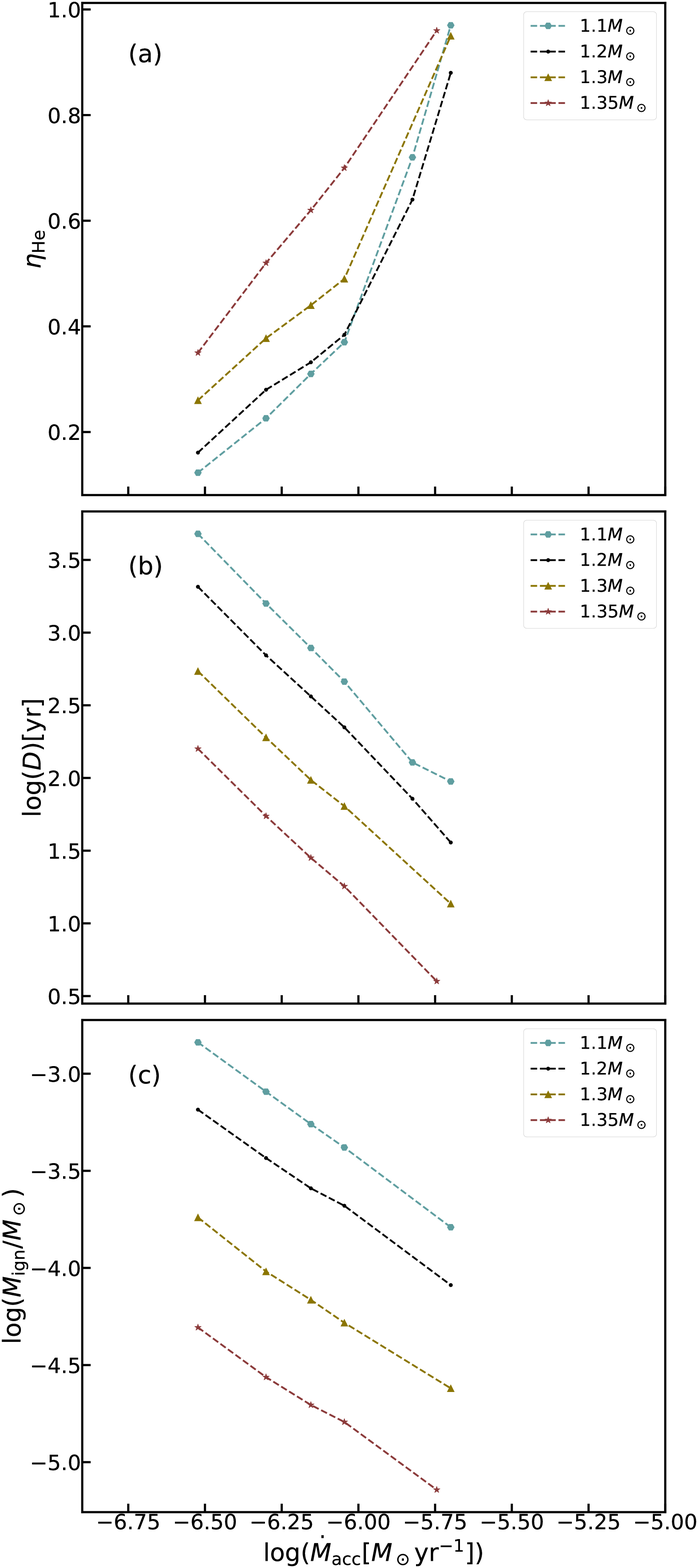,angle=0,width=9.5cm}
		\caption{Panel (a), (b) and (c) show the mass-retention efficiency $\eta_{\rm He}$, nova cycle duration $D$, and ignition mass $M_{\rm ign}$ changing with the accretion rate $\dot M_{\rm acc}$ for different ONe WD masses, respectively.}
	\end{center}
\end{figure}
Fig.\,3 shows the mass-retention efficiencies ($\eta_{\rm He}$), nova cycle durations ($D$) and ignition masses ($M_{\rm ign}$) changing with the accretion rates for different ONe WD masses.
Panel (a) presents the mass-retention efficiencies during He-shell flashes throughout the entire ($M_{\rm WD}$, $\dot M_{\rm acc}$) grid.
From this figure, we can see that the mass-retention efficiency increases with the mass-accretion rate for a given ONe WD mass.
This is because the degeneracy of He-shell is lower at higher accretion rates, resulting in weaker flashes of the He shell and thus less material to be blown away.
When the mass-accretion rate is close to the lower boundary of the helium steady burning region, the mass-retention efficiency would approach to 1, which means that almost all of the material remains on the surface of ONe WD.
It is also notable that the mass-retention efficiency increases with the ONe WD mass for a given accretion rate. This is because more massive WD have higher surface gravitation, resulting in nuclear burning rates much higher than wind mass loss rates. Therefore, we can infer that more material could accumulate on the surface of more massive ONe WDs for a given mass-accretion rate.

\begin{table}
	\centering
	\caption{Properties of He-shell flashes without mixing. }
	\begin{tabular}{ l  c  c ccc c  c  l }
		\toprule
		\hline \\
		$M_{\rm WD}$& $\dot M_{\rm acc}$&& $M_{\rm ign}$& $M_{\rm ej}$& log($D$)& $\eta_{\rm He}$	\\
		$(M_\odot)$&  $(M_\odot \rm yr^{-1})$&&     $(M_\odot)$&   $(M_\odot)$&  (yr) \\
		\hline \\
		1.1&        3e-7&&		       1.45e-3&    1.28e-3& 3.69&	0.12\\
		1.1&			5e-7&&  		   8.08e-4&    6.26e-4& 3.20&	0.23\\
		1.1&			7e-7&&  		   5.50e-4&	 3.76e-4& 2.90&	0.31\\
		1.1&			9e-7&&		       4.17e-4&    2.65e-4& 2.66&	0.37\\
		1.1&			1.5e-6&&		   2.06e-4&    8.37e-5& 2.11& 0.72\\
		\hline\\
		1.2&        3e-7&&		   6.46e-4&   5.47e-4& 3.31&	0.16\\
		1.2&			5e-7&&         3.51e-4&  2.54e-4& 2.84&	0.28\\
		1.2&			7e-7&&         2.56e-4&  1.72e-4& 2.57&	0.33\\
		1.2&			9e-7&&    	   2.00e-4&   1.26e-4& 2.35&	0.38\\
		1.2&		    2e-6&& 		8.16e-5&  1.86e-5& 1.56&	0.84\\
		\hline\\
		1.3&        3e-7&&			   1.66e-4&  1.23e-4& 2.74&	0.26 \\
		1.3&			5e-7&&			   9.72e-5&  6.26e-5& 2.28&	0.38\\
		1.3&			7e-7&	&		   6.60e-5&  3.59e-5& 1.99&	0.44\\
		1.3&			9e-7&	&		   5.53e-5&  2.72e-5& 1.81&	0.49\\
		1.3&            2e-6&&             2.21e-5&  1.95e-6& 1.14&	0.95\\	
		\hline\\
		1.35&       3e-7&&		   4.80e-5&      3.06e-5& 2.21 &	0.35\\
		1.35&			5e-7&&		   2.70e-5&	 1.26e-5& 1.74&	0.52\\
		1.35&			7e-7&	&		   1.96e-5&      7.62e-6& 1.45&	0.62\\
		1.35&			9e-7&	&		   1.59e-5&      4.87e-6& 1.26&	0.70\\
		1.35&		   1.8e-6&&		   8.39e-6&      7.28e-8& 0.60&	0.96\\
		
		\hline
	\end{tabular}
\end{table}

The cycle duration is an important observable feature for recurrent novae, which is defined as the time interval between two successive He-shell flashes (e.g., \citealt{2015MNRAS.446.1924H, 2016ApJ...819..168H}).
Panel (b) shows the nova cycle durations for different mass-accretion rates and ONe WD masses.
Obviously, the duration decreases with the mass-accretion rate for a given ONe WD mass, and also decreases with the WD mass for a given accretion rate.
This is because the He-shell mass required for nova outburst is smaller for massive ONe WDs due to their higher surface gravitation, as shown in panel (c).
For a given ONe WD, the accretion phase is shorter since the higher mass-accretion rates (e.g., \citealt{1982ApJ...257..312P}). 
Therefore, we can expect that the recurrent phenomena of He-shell flashes are more frequently to occur on massive ONe WDs with higher accretion rates.
\cite{2016ApJ...819..168H} investigate the He-accreting process of CO WDs, and suggested that the nova cycle duration decreases gradually if the He-shell has experienced enough flashes.
The reason is that the mass of WD is increased after multicycle flashes, and the WD is heated by the He-shell flashes, resulting in more frequent nova outbursts and weaker He-shell flashes.
Thus, we speculate that He-accreting ONe WDs can increase their mass after experiencing He-shell flashes, which may lead to  AIC events when their mass is close to $M_{\rm Ch}$, as studied in previous (e.g., \citealt{2017ApJ...843..151B, 2018RAA....18...36W}). 

The ignition mass $M_{\rm ign}$ is defined as the minimum helium shell mass that can trigger dynamical burning.
Panel (c) represents the He-shell ignition masses changing with ONe WD masses and mass-accretion rates.
From this figure, we can see that the ignition mass decreases with ONe WD mass and mass-accretion rate.
This is because the critical pressure required for the ignition of He-shell is a function of ONe WD mass and ignition mass, in which ignition mass varies inversely with ONe WD mass (see \citealt{1989clno.conf...39S}).
For a given ONe WD mass, a higher mass-accretion rate leads to a higher temperature growth rate of the accreted shell, resulting in a lower ignition mass (e.g., \citealt{1982ApJ...257..312P}).
Tab.\,1 displays the properties of He-shell flashes, including the ONe WD mass $M_{\rm WD}$, the mass-accretion rate $\dot M_{\rm acc}$,
the ignited mass $M_{\rm ign}$, the ejected mass $M_{\rm ej}$, the nova cycle duration $D$ and the mass-retention efficiency $\eta_{\rm He}$.
From this table, we can infer that mass-retention efficiency decreases with the ignition mass for a given ONe WD mass.
For example, in the case of a 1.35\,$M_\odot$ ONe WD, there is almost no mass loss when the ignition mass is $8.39 \times 10^{-6}M_\odot$, but almost all accretion material is lost when the ignition mass is $4.80 \times 10^{-5}\,M_\odot$.
If the ignition mass is smaller than a critical value, the mass-retention efficiency would be close to 1, which means that no mass loss would occur (see also \citealt{1989ApJ...340..509K}).

\begin{table}
	\centering
	\caption{Mass fractions of nuclear products during a nova cycle.}
	\begin{tabular}{ l  c  c cccc c  c  l }
		\toprule
		\hline \\
		Set & $M_{\rm WD}$ $(M_\odot)$&$\dot M_{\rm acc}$ $(M_\odot \rm yr^{-1})$&C&O&Ne&Mg&Si \\
		\hline \\
		1&  1.20&3e-7&4.10e-01&1.66e-01&7.84e-02&3.36e-01&7.60e-03\\
		2&  1.20&5e-7&4.42e-01&1.95e-01&8.65e-02&2.73e-01&6.00e-03\\
		3&  1.20&7e-7&4.56e-01&2.14e-01&8.83e-02&2.36e-01&4.99e-03\\
		4&  1.20&9e-7&4.55e-01&2.53e-01&9.07e-02&1.96e-01&4.60e-03\\
		5&  1.35&7e-7 & 2.22e-01&6.08e-03&9.76e-03&5.41e-01&2.21e-01\\
		\hline
	\end{tabular}
\end{table}

\begin{figure}
	\begin{center}
		\epsfig{file=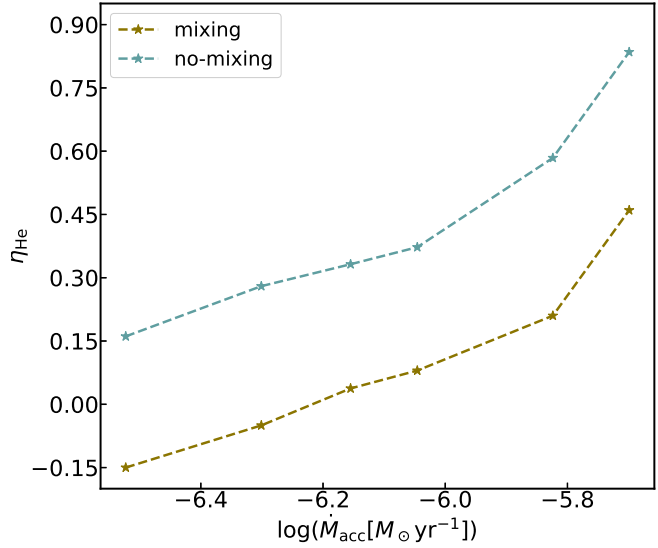,angle=0,width=9.5cm}
		\caption{Influence of mixing on the mass-retention efficiencies of a 1.2\,$M_\odot$ ONe WD. }
	\end{center}
\end{figure}
\begin{figure}
	\begin{center}
		\epsfig{file = 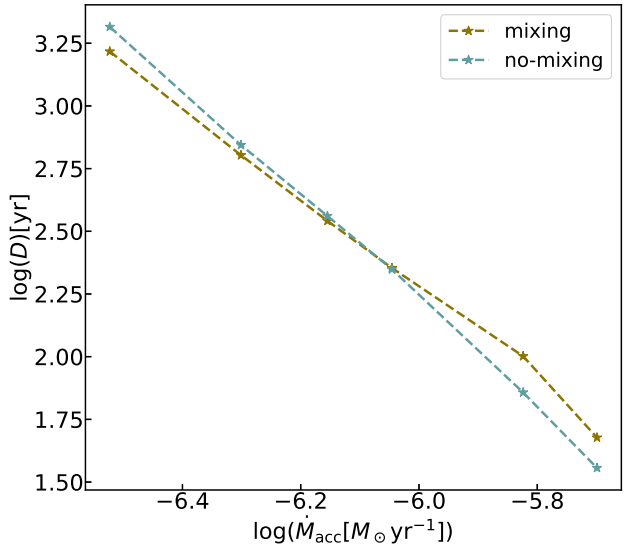,angle=0,width=9.5cm}
		\caption{Influence of mixing on the nova cycle durations of a 1.2\,$M_\odot$ ONe WD.}
	\end{center}
\end{figure}

\begin{figure}
	\begin{center}
		\epsfig{file = 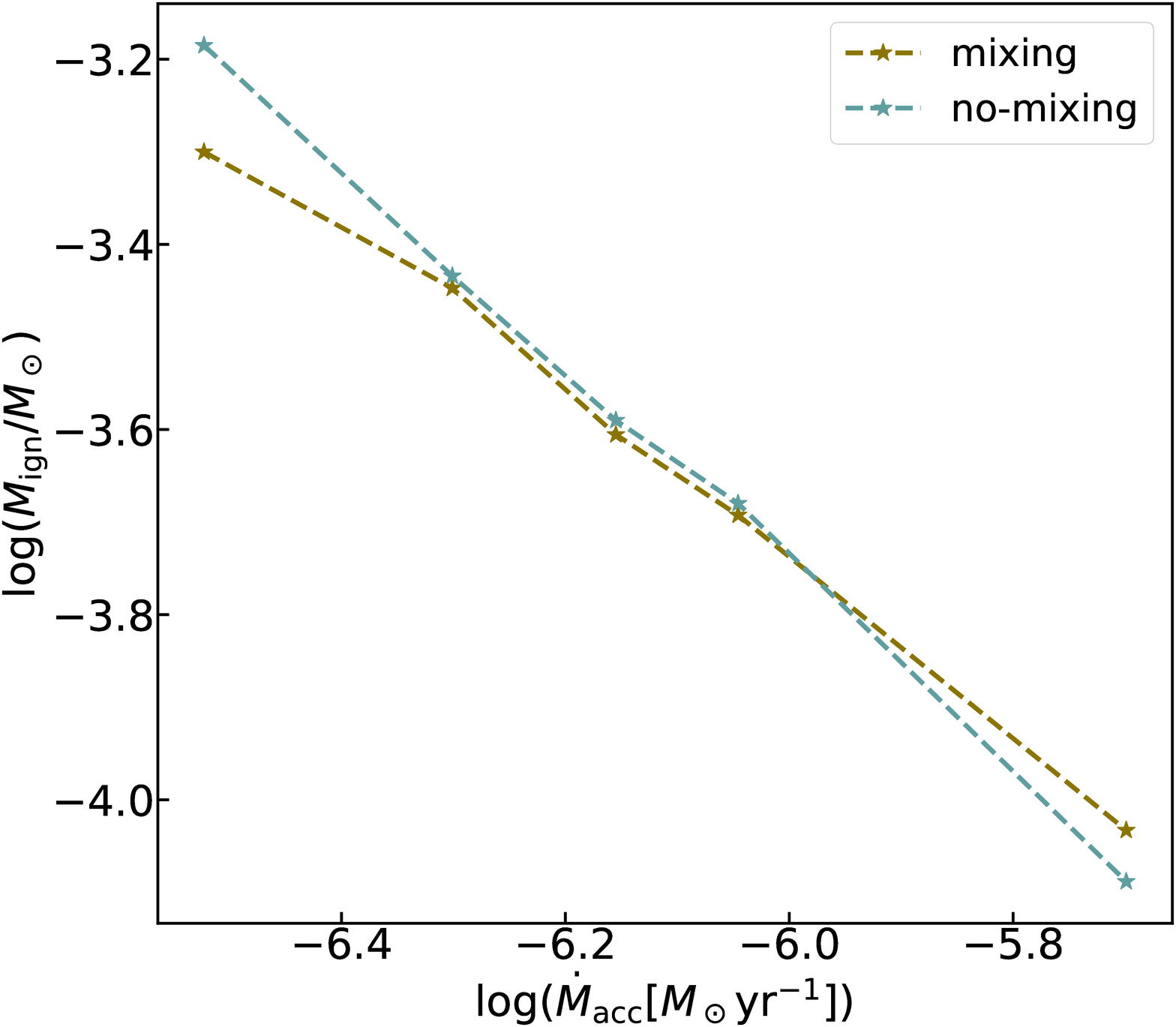,angle=0,width=9.5cm}
		\caption{Influence of mixing on the ignition masses of a 1.2\,$M_\odot$ ONe WD. }
	\end{center}
\end{figure}

According to the results of \cite{2005ApJ...623..398Y}, a nova burst occurs only when the cooling timescale ($\tau_{\rm cool}$) is longer than the accretion timescale ($\tau_{\rm acc}$).
Obviously, the value of $\tau_{\rm acc}$ is much lower for a higher accretion rate, resulting in weaker He-shell flashes and less ejected mass (see also \citealt{2012BASI...40..419S}).
On the contrary, a lower accretion rate results in a stronger degenerate helium envelope since long cooling time. If the conditions of He-shell flashes can be reached by gradually accumulating He-rich materials, a stronger burst will be triggered and blows away more material.
However, we found that the temperature of He-shell drops quickly and it is hard to reach the conditions of He-shell flashes for very low accretion rates.

The chemical abundances (mass fraction) of nuclear products during a nova cycle are summarized in Tab.\,2. It is apparent that the elements produced during the He-shell flashes are sensitive to the ONe WDs masses and accretion rates. We found that C, O and Ne increase with mass-accretion rates while Mg and Si decrease. This is because higher mass-accretion rates result in higher nuclear burning rates and weaker He-shell flashes. Thus, more He-rich material could be burn into C, O and then Ne. On the other hand, as the peak temperature is higher for low mass-accretion rates, more heavier elements, such as Mg and Si are produced. In addition, C, O and Ne decrease with WD masses while Mg and Si increase. This is due to the higher peak temperature caused by the stronger surface gravity acceleration for a massive WD. The trend of this result is consistent with \cite{2018ApJ...863..125K}.
\subsection{Influence of mixing process}
The mass-retention efficiencies provided in this work are likely the lower limit. We adopt a super-Eddington wind during the He-shell flashes, in which the material is assumed to be ejected if the escape speed is reached.
In addition, if the rotation of ONe WD is considered, the He-shell flash would be much weaker since the weaker effective gravity and the material of WD core by rotational mixing at the base of the He-shell, which may lead to larger mass-retention efficiencies (see \citealt{2004A&A...425..217Y}).
Moreover, the mass of ONe WD can exceed $M_{\rm Ch}$ when the rotation effects are taken into account (e.g., \citealt{2004A&A...419..623Y, 2009ApJ...702..686C, 2011ApJ...730L..34J, 2012ApJ...744...69H, 2014MNRAS.445.2340W}).

\begin{table}
	\centering
	\caption{Properties of He-shell flashes with mixing, in which $M_{\rm WD}$ = 1.2\,$M_\odot$. }
	\begin{tabular}{ l  c  c ccc c  c  l }
		\toprule
		\hline \\
		$\dot M_{\rm acc}$&& $M_{\rm ign}$& $M_{\rm ej}$& log($D$)& $\eta_{\rm He}$	\\
		$(M_\odot \rm yr^{-1})$&&     $(M_\odot)$&   $(M_\odot)$&  (yr) \\
		
		\hline\\
		3e-7&&		   6.20e-4&   4.91e-4& 3.22&	-0.15\\
		5e-7&&         3.49e-4&  2.50e-4& 2.80&	-0.05\\
		7e-7&&         2.44e-4&  1.14e-4& 2.54&	0.04\\
		9e-7&&    	   2.06e-4&  8.51e-5& 2.35&	0.08\\
		1.5e-6&&	   1.39e-4&	 4.13e-5& 2.02& 0.21\\
		2e-6&&		9.27e-5& 7.54e-6& 1.68&	0.46	\\	
		
		\hline
	\end{tabular}
\end{table}

A number of Ne enriched novae have been observed, e.g., V693 CrA 1981, V1370 Aql 1982, QU Vul 1984, V1668 Cyg 1978, V838 Her 1991, 1974 Cyg 1992 (e.g., \citealt{1998PASP..110....3G, 1999ApJ...523..409W, 2013ApJ...762..105D}).
An important feature of Ne novae is the significant enrichments of O, Ne, Mg and Al detected in the ejection (e.g., \citealt{1994A&A...291..869A, 1998ApJ...494..680J, 1998PASP..110....3G, 1999ApJ...523..409W}).
\cite{1998ApJ...494..680J} investigated the elemental abundance in the ejection of Ne novae, and found that the considered mixing model may be a suitable way to explain the observed element abundance.
During the mass-accretion process, the accreted envelope may be mixed with the material in the underlying WD core.
The efficiency of mixing on accreting WDs depends on which diffusion physics are included,
e.g., if magnetic torques are included, mixing during accretion tends to be minimal
(e.g., \citealt{2004A&A...419..623Y, 2017A&A...602A..55N}).
However, the mixing mechanism is not completely clear.
There are some hypotheses for the mixing mechanism, such as diffusion, shear mixing, convective overshooting and Kelvin-Helmholtz instabilities (e.g., \citealt{1984ApJ...281..367P, 1987ApJ...321..386K, 1991ApJ...375L..27I, 1997ApJ...475..754G, 1998A&A...337..379K, 2016A&A...595A..28C}).

Here, we provide a simplified mixing case without including the physical mechanism of the mixing process, as described in \cite{1995ApJ...448..807P}. We assume that the accreted material from the companion is mixed by a given element fraction of the underlying ONe WD core. We set the mixing fraction to be 0.5, which has been suggested as a representative value (e.g., \citealt{1994ApJ...425..797L}). In Tab.\,3, we provide the properties of He-shell flashes for the mixing cases.
Fig.\,4 shows the influence of mixing on the mass-retention efficiencies of a 1.2\,$M_\odot$ ONe WD. It is obvious that the mass-retention efficiency in the mixing case is lower than that without considering mixing for a given mass-accretion rate.
Especially for the cases with low mass-accretion rates (e.g., $\dot M_{\rm acc}$ = $3 \times 10^{-7}$\,$M_\odot \rm yr^{-1}$), the value of $\eta_{\rm He}$ is negative, which means that the ONe WD mass is decreasing. This is because the core material is dredged up from the underlying ONe WD during mixing process, and thus ONe WD is losing mass as a result of continued outbursts. 
Meanwhile, even in the case of higher accretion rates, the mass of ONe WD under mixing conditions is more difficult to grow than the cases without mixing.
This result indicates that the consideration of the mixing process may also affect the predicted birthrates of AIC events. Therefore, we suggest that the mixing process is important for the study of the final fate of He-accreting ONe WDs.
In addition, Figs 5-6 present that the influence of mixing on the nova cycle duration and ignition mass of a 1.2\,$M_\odot$ ONe WD. These two figures indicate that the mixing process has slight influence on the nova cycle duration and ignition mass. 
This is because the nova cycle duration and ignition mass are more sensitive to the temperature and density of the accreted material, but not very sensitive to their composition.

\section{Discussion and Conclusions} \label{6. DISCUSSION}

It has been suggested that AIC events are about 5 mag fainter than typical normal SNe Ia, and can only last from a few days to a week,
so that AIC is difficult to be identified in the observations.
According to \cite{2013ApJ...762L..17P}, radio sources that last for a few months may be caused by AIC events.
In addition, it is possible that a small amount of $\prescript{56}{}{\mathbf{Ni}}$ is synthesized in the AIC process
(e.g., \citealt{2006ApJ...644.1063D, 2010MNRAS.409..846D}). 
These characteristics may be used to identify AIC events.
Tab.\,2 shows that more Mg and Si can be produced in the He-shell flashes for massive ONe WDs.
We expect that the surface of the WDs would be covered with a Mg/Si-rich shell 
if the WDs masses increase to approach $M_{\rm Ch}$ by accreting He-rich material.
Therefore, the enrichments of Mg/Si may be detectable in AIC events formed by ONe WDs accreting He-rich material.

In addition, it is difficult to accurately distinguish whether the accretor is a CO WD or an ONe WD if the accretion rate is significantly large, as there is no obvious mixing process or significant enrichment elemental abundance of WD in the ejection.
As mentioned above, V445 Pup is the only He nova that has been identified so far.
The WD mass is suggested to be $\geq$ 1.35\,$M_\odot$ and half of the accreted material could remain on its surface (see \citealt{2008ApJ...684.1366K}).
\cite{2008ApJ...684.1366K} suggested that the WD of V445 Pup is a CO WD but not an ONe WD since no signatures of neon enhancement were detected, and argued that V445 Pup is a strong candidate of SNe Ia (see also \citealt{2005ASPC..330..451W, 2018RAA....18...49W}).
However, the carbon features originated from WD core or helium burning are still not determined, and it is still unclear whether there is a mixing process or not during mass-accreting process (e.g., \citealt{2004ApJ...613L.129K}).
Further detailed researches are needed in future studies to provide more precise nature of the WD in V445 Pup.

By using the stellar evolution code MESA, we investigated the long-term evolution of accreting ONe WDs that undergo multicycle He-shell flashes with initial masses ranging from 1.1 to 1.35\,$M_\odot$ and various accretion rates in the range of $3 \times 10^{-7}$$-$$2 \times 10^{-6}$ ${M}_\odot \rm yr^{-1}$.
We found that the mass-retention efficiency both increases with the ONe WD mass and the mass-accretion rate.
More massive ONe WDs or higher mass-accretion rates lead to shorter nova cycle durations, which indicates that the ONe WD mass grows faster for the cases with more massive initial ONe WDs and higher mass-accretion rates. 
Meanwhile, we also found that the ignition mass both decreases with the ONe WD mass and the mass-accretion rate.
We also present the nuclear products in different accretion scenarios.
In addition, for a given ONe WD mass and mass-accretion rate,
the mass-retention efficiency would be lower after considering the mixing effects. 
In order to put more constraints on the process of He-shell flashes, we hope that more He novae and Ne novae can be identified in the observations.

\begin{acknowledgements}
We thank the anonymous referee for valuable comments that can help to improve the paper. BW is supported by the National Natural Science Foundation of China (Nos 11873085, 11673059 and 11521303),
the Chinese Academy of Sciences (No QYZDB-SSW-SYS001),
and the Yunnan Province (Nos 2018FB005 and 2019FJ001).
DDL is supported by the National Natural Science Foundation of China (No. 11903075) and the Western Light Youth Project of Chinese Academy of Sciences.

\end{acknowledgements}

\bibliography{1bib}
\bibliographystyle{raa}

\end{document}